# Information Security Strategy in Organisations: Review, Discussion and Future Research Directions

**Craig A. Horne**
Department of Computing and Information Systems
The University of Melbourne
Victoria, Australia
Email: chorne@student.unimelb.edu.au

**Atif Ahmad**
Department of Computing and Information Systems
The University of Melbourne
Victoria, Australia
Email: atif@unimelb.edu.au

**Sean B. Maynard**
Department of Computing and Information Systems
The University of Melbourne
Victoria, Australia
Email: sean.maynard@unimelb.edu.au

## Abstract

Dependence on information, including for some of the world's largest organisations such as governments and multi-national corporations, has grown rapidly in recent years. However, reports of information security breaches and their associated consequences continue to indicate that attacks are still escalating on organisations when conducting these information-based activities. Clearly, more research is needed to better understand how organisations should formulate strategy to secure their information. Through a thematic review of academic security literature, we (1) analyse the antecedent conditions that motivate the potential adoption of a comprehensive information security strategy, (2) the current perspectives of strategy and (3) the yields and benefits that could be enjoyed post-adoption. Our contributions include a definition of information security strategy. We argue for a paradigm shift to extend from internally-focussed protection of organisation-wide information towards a strategic view that considers the inter-organisational level. Our findings are then used to suggest future research directions.

## 1 INTRODUCTION

Information resources play a critical role in sustaining business success by driving innovation and opportunities for the development of competitive advantage. As such, preservation of the confidentiality, integrity and availability of these information resources is a significant imperative for organisations, as is the need for a viable information security strategy in organisations (ISSiO) to facilitate information transfer at an inter-organisational level.

The aim of this paper is to identify a strategic approach to securing information resources for the benefit of those decision-makers accountable for driving strategic-level organisational security and ultimately organisational success. The scope of the research is to examine the conceptual construct of ISSiO. In particular, the authors of this paper are motivated by calls from other information systems researchers for the development of a comprehensive security strategic framework (Baskerville et al. 2014), and for future research into the role that boards of directors may play in information security practices (McFadzean et al. 2006).

Significantly, some of the world's largest organisations, including governments and multi-national corporations, have quite publicly suffered security incidents. By broadly reviewing the extant literature, a perspective will be established that can support the development of a comprehensive ISSiO which could be generalisable to all organisations. This paper is a critical literature review on the topic of ISSiO. Papers from various researchers were analysed and evaluated before being compared for depth of understanding and conclusions drawn. The paper commentary is explicative, interpretative and centres on the determination of the theory of ISSiO.

The paper continues in four major sections. Initially we introduce ISSiO, discuss its origins and existing definitions whilst expanding on some of its more central properties. Second, we review the construct space of ISSiO to understand prior research on how ISSiO is conceptualised, the level of analysis from which ISSiO is approached and contend with propositions for measuring the distinct elements of an ISSiO. Third, we review the nomological network space to assess the environmental antecedents, conceptual elements, and possible yields from an ISSiO. Finally, we draw conclusions, construct a definition, consider limitations and provide suggestions for future research to advance our understanding of information security strategy.

## 2   DEFINING INFORMATION SECURITY STRATEGY

Definitions of ISSiO are infrequent in the information systems literature so in this section, in an indulgent departure from convention, the discussion is largely author-centric rather than concept-centric.

Information security strategy is defined by Beebe and Rao (2010, pg. 330) as "*the pattern or plan that integrates the organisation's major IS security goals, policies, and action sequences into a cohesive whole*". These authors believe ISSiO is a documented plan which matches an assessment of external cyber threats with a financially-informed set of internal countermeasures, including the required supporting policies and procedures. Strategy is seen as the means to influence an organisation's environment through the careful selection of internal controls.

Park and Ruighaver (2008, pg. 27) define information security strategy as:

> "an art of deciding how to best utilize what appropriate defensive information security technologies and measures, and of deploying and applying them in a coordinated way to defence (*sic*) organisation's information infrastructure(s) against internal and external threats by offering confidentiality, integrity and availability at the expense of least efforts and costs while to be effective".

These authors believe ISSiO has been developed from the military literature and therefore tends to be focussed more on how to deploy strategies than focus on what goals the organisation is trying achieve. In terms of attempting to classify ISSiO, their analysis of earlier literature leads them to the conclusion that ISSiO balances three dimensions which are time, space and the decision-making process.

Ahmad et al. (2014b) and Park and Ruighaver (2008) believe ISSiO can be used to incrementally improve the quality of the information security program, however there must be a strong link from the ISSiO to the organisational strategic plan to support it. ISSiO is necessary to prevent threats to an organisation's information. ISSiO can take the form of one of a number of areas which include deterrence, prevention, surveillance, detection, response, deception, perimeter defence, compartmentalisation and layering. Senior business sponsorship of the security function is also required.

Hong et al. (2003) do not define ISSiO per se but assert that it is a function of policy orientation, risk management orientation, control and auditing orientation, management systems orientation and contingency management. Contingency management is assessed by the authors as a function of the organisational environment, management and technology. Sveen et al. (2009) contend that an ISSiO is like any other business strategy: it is the process of building up resources. By simply explaining what an ISSiO is, Sveen et al. (2009) describe the construct but have not provided a formal definition. Their insights are still useful however in building up our cumulative understanding.

These definitions give an insight into the difficulties with achieving unanimity on defining ISSiO. Using conceptualisation of ISSiO as an example, Beebe and Rao (2010) explain it is a plan, Sveen et al. (2009) assert it is a process and conceptualisations from Park and Ruighaver (2008), Ahmad et al. (2014b) and Hong et al. (2003) do not fit within either of these. There are many other researchers who have used the term 'information security strategy' in their literature however they have not provided an explicit definition.

### 2.1   Information Security Strategy: Plan or Process?

There are two main conceptualisations espoused by organisational scholars when describing ISSiO. These include (1) a static plan, described as an artefact to be shared amongst stakeholders (Beebe and Rao 2010; Bowen et al. 2006; Von Solms and Von Solms 2004), and (2) a dynamic process, to be followed by stakeholders concerned with protecting organisational information (Booker 2006; Brotby et al. 2006; Flores et al. 2014; McFadzean et al. 2006; Sveen et al. 2009; Van Niekerk and Von Solms 2010). A profound comprehension of these interpretations will shed light on how to apply them in ISSiO research.

Some information systems researchers view ISSiO as a static plan; a central artefact to be developed that describes the linkages between various organisational concepts such as goals, policies and action sequences (Baskerville and Dhillon 2008; Beebe and Rao 2010). In a process orientation, ISSiO involves using a strategy-setting process, whilst incorporating the organisational information systems security goals, such as regulatory compliance, as input. This strategy-setting process can group actions taken according to either the end product ultimately derived such as a strategic security plan, or the processes required such as aligning ISSiO with organisational strategy (Baskerville and Dhillon 2008). Finally some information systems scholars do not conceptualise ISSiO at all or characterise it in abstract terms only (Hong et al. 2003; Park and Ruighaver 2008).

## 3 INFORMATION SECURITY STRATEGY IN INFORMATION SYSTEMS RESEARCH

A number of information systems researchers have made individual contributions towards understanding ISSiO from various perspectives. The focus of these researchers was to address problems including adequate support for organisational strategic vision, information systems-business cohesiveness and coordination of information security efforts. However, a complete and methodical evaluation of ISSiO within the information systems literature has not been accomplished. Therefore our research seeks to firstly examine what information systems researchers have analysed about the ISSiO construct and secondly the ISSiO nomological network describing its various elements. The ISSiO construct denotes the theoretical domain of ISSiO, specifically how it is conceptualised, at what levels of analysis it can be stratified, and measurement proposals to ensure unit specificity. The ISSiO nomological network refers to our understanding of ISSiO phenomena in the information systems domain, captured through the completion of a thematic analysis.

### 3.1 Literature Review

Our initial search for information security strategy was for manifestations of it in peer-reviewed information systems journals and selected conference proceedings, found through searching institutional repositories, Google Scholar and A* information systems journals. Our search consisted of articles that included the complete search string "information security strategy" in English. We searched backwards to discover prior articles and forwards for articles that cited seminal articles (Webster and Watson 2002). We did not restrict the search based on article age or grade of journal, preferring instead to examine each artefact found for nuances, no matter how small, which could shed light on our evolving understanding of the concept. We also included papers that referred to "information security" but included the word strategies (plural) instead, to facilitate an investigation for example into whether use of the singular 'strategy' or plural 'strategies' could indicate a shift in level of analysis within an organisation. Finally, we included papers that centred on information security but discussed an implicit aspect of strategy. Note that 'organisation' is a term used to denote private companies, public governments, not-for-profit societies and educational institutions.

We included an international standard on information security, as we thought this could have important implications for motivating the use of an ISSiO; however we did not include any practice-oriented literature such as vendor white papers due to issues with accessibility and peer-review process. Out of the results, 45 papers were deemed of interest.

We then examined each paper to explore how ISSiO relates to the article's core paradigm. The following four classifications stratify how central ISSiO is to each paper and is adapted from Roberts et al. (2012):

1. *Implicit use of the term*. Information security forms the paper's central theme and strategy is implicit only. Information security strategy does not form the central argument of the paper, e.g. (Van Niekerk and Von Solms 2010).

2. *Provides conceptual support*. Papers use information security strategy to support the development of their concepts, e.g. (Flores et al. 2014).

3. *Used in the research question or hypothesis*. Papers use information security strategy explicitly in their findings or analysis, e.g. (Posthumus and von Solms 2004).

4. *Forms the conceptual base for the paper.* These papers are entirely consumed with the discussion of information security strategy, e.g. (Baskerville and Dhillon 2008).

In summary, 35 percent of articles that were collected implied some aspect of ISSiO when discussing information security. 27 percent of articles provided theoretical or conceptual support for developing the logic of ISSiO. 18 percent of articles used ISSiO in some part of their hypothesis, research question or proposition. One fifth of articles were focussed purely on discovery of aspects relating to ISSiO. In the next section, we discuss the role of ISSiO in information systems research in more detail.

## 3.2 The Information Security Strategy Construct

From the previous sections, it could be perceived that ISSiO has not been widely developed in the information systems literature so a more profound analysis is warranted. The following sections discuss in more detail ISSiO's (1) conceptualisation, (2) levels of analysis and (3) measurement domain.

### 3.2.1 Conceptualisation

We examined what researchers understood the main conceptual context for the ISSiO construct was. The three groups used for this construct are firstly as a plan, secondly as a process, and thirdly neither of these.

Table 1 presents some conceptualisations (i.e. plans, processes, or neither conceptualisation) and the role of ISSiO in the information systems literature. Out of the 45 articles that were examined, 20 percent (9 papers) used ISSiO as the core of the entire article. 78 percent (35 papers) gave neither explicit conceptualisation of ISSiO. In terms of patterns, when ISSiO is used in the research question (row 3) or forms the theoretical basis for the paper (row4), it becomes apparent that ISSiO is largely viewed by information systems authors as neither plan nor process.

|  | Plan | Process | Neither Plan nor Process | Total |
| --- | --- | --- | --- | --- |
| 1. Implicit use of the term | 1 | 1 | 14 | 16 |
| 2. Provides conceptual support | 1 | 3 | 8 | 12 |
| 3. Used in research question or hypothesis | 0 | 1 | 7 | 8 |
| 4. Forms theoretical basis for paper | 1 | 2 | 6 | 9 |
| Total | 3 | 7 | 35 | 45 |

*Table 1. Information Security Strategy Conceptualisations and Role in Information Systems Research*

### 3.2.2 Levels of analysis

For the purposes of clarification, in this paper a group is a set of individuals who are responsible for some aspect of security within an organisation. Also, in this section where a paper discusses aspects of responsibility for the application of ISSiO at two different levels, the higher of the two was recorded for the purpose of this analysis. This is because the higher level is seen to be more complex, with greater relationship interdependencies.

Table 2 shows that while ISSiO is acknowledged to be a multilevel construct, researchers (with only 3 from 45 papers, or 7 percent) do not typically characterise ISSiO from an individual perspective. A significant 60 percent (27 from 45 papers) of the information systems literature examined contend that ISSiO belongs at an organisational level. At an organisational or inter-organisational level, it is apparent (with 35 from 45 papers, or 78 percent) that scholars believe ISSiO is neither plan or process.

|  | Plan | Process | Neither Plan nor Process | Total |
| --- | --- | --- | --- | --- |
| 1. Individual | 0 | 0 | 3 | 3 |
| 2. Group | 0 | 1 | 7 | 8 |
| 3. Organisation | 3 | 5 | 19 | 27 |
| 4. Inter-organisational | 0 | 1 | 6 | 7 |
| Total | 3 | 7 | 35 | 45 |

*Table 2. Information Security Strategy Conceptualisations and Levels of Analysis*

### 3.2.3   Measurement domain

When operationalising ISSiO, if conceptual elements cannot be measured, then their reliability cannot be known. There are eight papers in the information systems literature that use the term 'information security strategy' and expand the theoretical base of ISSiO. Of these, 75 percent (6 from 8 papers) contend that ISSiO exists at an organisational level. Half of these (4 from 8) papers hold that ISSiO is neither a plan nor a process.

A number of these papers confusingly use the word 'measure' as an abbreviation for 'countermeasure', which is a control installed to mitigate the risk arising from a threat to an asset (Ahmad et al. 2014b; Beebe and Rao 2009; Park and Ruighaver 2008). Two papers contained no mention of 'measure' at all (Hong et al. 2003; Kayworth and Whitten 2010).

Of the three papers that addressed the measurement of some aspect of ISSiO, the main areas which were measurable included risk management, goal achievement and quality. Risk management can be measured by efficacy, efficiency or effectiveness (Baskerville and Dhillon 2008), time can be a primary measure of risk (Baskerville et al. 2014) or alternatively an examination of a finite set of risk-reducing countermeasures can be measured (Beebe and Rao 2010). Goal achievement is measured by the activities undertaken to achieve those goals (Baskerville and Dhillon 2008). Quality improvement can be gained through the measuring of routine security tasks (Baskerville et al. 2014).

## 3.3   The Information Security Strategy Nomological Network

In this section we undertake a thematic analysis within the information systems literature to conceptualise ISSiO at various levels within an organisation and develop a nomological network map to explain the construct and its interrelationships. Thematic analysis is a common technique that has been used by other researchers to examine large bodies of work within the information systems literature (Leidner and Kayworth 2006; Roberts et al. 2012). Thematic analysis is the process of conducting a qualitative content analysis on the literature of interest then listing meritorious ideas from each article before organising them related groups (Cline and Jensen 2004). To conduct the thematic analysis, we firstly analysed 45 papers for their interpretation of ISSiO and then grouped key constructs according to similarities of themes. This resulted in three distinct themes emerging from the analysis, which were antecedents, constituents and possible yields.

Antecedents are the precursor conditions that might prompt an organisation to consider the use of an ISSiO. Examples include governments with top secret files, pharmaceutical companies conducting extended clinical new drug trials and banks facilitating online trading. Constituents are the elements that make up the core of an ISSiO, to be adopted by an organisation seeking to protect its information. Examples include risk management process to understand persistent common threats, security auditing to satisfy external regulators and governance activities to align organisational efforts. Yields are the benefits that can be enjoyed after successfully adopting ISSiO. Examples include the confidentiality, integrity and availability of information, protection of competitive advantage and brand protection and trust.

Based on the thematic analysis and discussion in preceding sections, a logical grouping of the conceptual elements of ISSiO can be elicited from the literature and is shown in Figure 1.

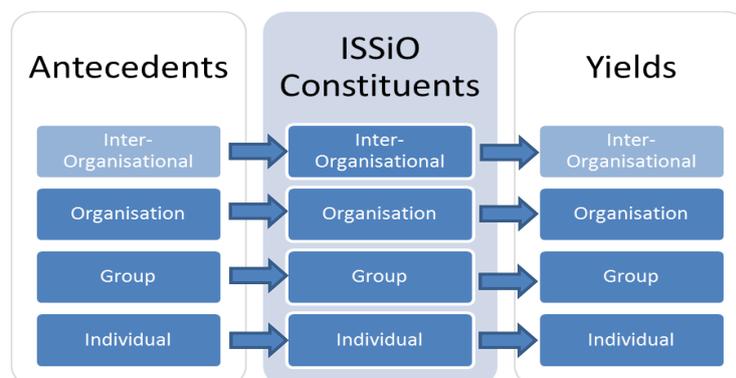

*Figure 1: Thematic Map of Information Security Strategy in Organisations in IS Research*

The sections below discuss these themes in more detail. Our assessment is focussed on conceptual elements that can be contributed from each paper and an overall understanding of what each author believes ISSiO is.

### 3.3.1 Antecedents

Antecedents are the precursor conditions necessary to prompt the use of ISSiO and emerged as a theme in the information systems literature after conducting a thematic analysis, as described in the previous section.

At an individual level, there did not seem to be any antecedents apparent in the literature. It is impossible to make an exhaustive claim about this but perhaps this is an area that warrants further attention from researchers.

At a group level, one ISSiO antecedent is the requirement for global ubiquitous information availability and the necessity to distil incomprehensible threat intelligence complexity and volume in a timely fashion (Booker 2006).

At an organisational level, there are few antecedents for ISSiO apparent in the literature but gathering intelligence about the external environment was one. An organisation's information security strategic posture involves a dependence on the external threat environment, not the continued successful achievement of organisational goals. The increasing complexity and sophistication of dynamic, targeted attacks over time naturally causes a general shift in posture balance from preventative towards a more response-oriented approach (Baskerville et al. 2014). Organisational ownership of information assets of value is also a key driver towards the adoption of ISSiO (Kelly 1999).

At the inter-organisational level, an ISSiO must take into consideration an organisation's regulatory compliance burden (Banker et al. 2010; Kayworth and Whitten 2010; Tutton 2010). This regulatory compliance-driven approach however only forms part of a holistic approach to security (Anderson and Choobineh 2008). Regulatory and legal compliance along with adoption of standards and best practices is also required (Posthumus and von Solms 2004). Examination of the industry in which the organisation competes and sufficient knowledge of industrial and economic considerations of an organisation's competitive landscape are also required (Baets 1992). The existence of a strategic information systems plan is notable, as it dictates the formulation of the information security policy by providing essential details of the business context or competitive landscape (Doherty and Fulford 2006). Failure of political pressure and economic sanctions are important preconditions that may motivate the commencement of information warfare (Baskerville 2010). ISSiO is primarily based on prevention of incidents arising from advanced persistent threats (APT) using technical controls against external threats that are seen to be increasingly more frequent, novel and costly (Beebe and Rao 2009). Environmental and organisational conditions, managerial understanding and actions, quality improvement initiatives and organisational achievement lead to use of ISSiO (Cline and Jensen 2004). Regulatory, political and legal compliance plus adoption of standards and best practices motivate the use of ISSiO (Kim et al. 2012; Posthumus and von Solms 2004). Standards exist which detail management of information security which in turn could assist with ISSiO development (Brotby et al. 2006; ISO/IEC 2013).

### 3.3.2 Constituents

Constituents are the central conceptual elements of ISSiO and emerged as a theme in the information systems literature after conducting a thematic analysis, as described in Section 3.3.

*Individual level*

This section seeks to explore what role an individual has in contributing towards the overall success of the strategic use of information security. At an individual level, there were no constituent elements of ISSiO however this is unusual because it is widely accepted that overall security depends on the weakest link which is typically the individual. This may represent an opportunity for further research.

*Group level*

This section examines the IS literature to discover the dynamics of groups working to support the strategic use of information security. At a group level, the constituent components of the ISSiO construct are varied and numerous. One is the identification and protection of knowledge assets, which can be resources forming a competitive advantage and can be either held in the human brain or in organisational documents, routines, procedures and practices. Knowledge leakage is a security incident which can temporarily affect an organisation's competitive advantage and affect its reputation, revenue streams, remediation costs and productivity. Mitigation or protection of knowledge is achieved through initial classification of information assets, then compartmentalisation, development of technical solutions, policies, procedures, culture and legal support (Ahmad et al. 2014a). ISSiO should guide the overall security budget for an organisation, to enable the security staff group and their management to fund and implement security resources that optimise security outcomes based on expense versus benefits (Anderson and Choobineh 2008). ISSiO includes the examination of stratified responsibility within an organisation that cohesively achieves overall information systems security. Decisions made by one layer of responsible agents affect decisions made by agents in other layers and their communication is vital. ISSiO success depends on action taken by responsible agents rather than technological controls. Achievement of ISSiO allows alignment with policies and regulatory compliance efforts (Backhouse and Dhillon 1996).

An essential element of ISSiO is a mix of technical, formal and informal controls to ensure regulatory compliance, protect the IT infrastructure that the information resides on and deliver CIA to users (Beebe and Rao 2009; Posthumus and von Solms 2004; Sveen et al. 2009). Security education, training, awareness and constant monitoring are required to ensure employees can use controls (Taylor and Robinson 2014; Van Niekerk and Von Solms 2010). ISSiO includes the capability to respond to attacks effectively, which stems from supplementary forces creating a time buffer through the employment of defence-in-depth design to allow the responding forces enough to time to deploy to the breach from the central holding point (Burnburg 2003). Information systems solutions underpin business products and services and are therefore critical in maintaining an organisation's competitive advantage. An ISSiO must focus on how to maintain competitive advantage in the face of rapidly changing ICT infrastructures (Cegielski et al. 2013).

*Organisational level*

The organisational level is where most influence can be exerted internally to achieve success in supporting an externally-focussed strategic application of information security and deserves special attention in our examination of the IS literature. At an organisational level, ISSiO can be used to incrementally improve the quality of the information security program. There must a strong link from the ISSiO to the business strategic plan to support it. ISSiO is necessary to prevent threats to an organisation's information (Ahmad et al. 2014b). It supports incremental quality improvement, alignment with agency mission, and awareness and monitoring of external threats (Bowen et al. 2006; Johnson and Goetz 2007). ISSiO protects only the more valuable information assets in order to reduce expenditure. This is achieved through policies and communication structures, director-level sponsorship of security initiatives, measuring success and administering sanctions for security policy violations. Identity and access management is important to overall success as is security incident detection and response activities (Ahmad et al. 2012; Kelly 1999). Corporate knowledge assets can then be inventoried and values defined (Baets 1992).

If the labour involved with security functions is outsourced to other companies or individual contractors, then they need to equally adhere to the security policies and strategy adopted by the parent organisation (Baskerville et al. 2014). ISSiO can use SCP to introduce a deterrent option within the risk management section (Beebe and Rao 2010). It is centred in risk management, identifying controls to mitigate known threats (Da Veiga and Eloff 2007). Reducing risk lowers anticipated loss, which changes an organisation's security posture. Quantifying risk of anticipated loss requires recording of previous loss from security incidents (Ryan and Ryan 2006). Conceptual constituents also include regulatory compliance, teleworkers, organisational agility, business justification requirements, reactive quality improvement and community cloud initiatives (Booker 2006). The external environment places various demands on the organisation which changes to continue the achievement of the organisational objectives. The ISSiO is contingent on the environment when changing to maintain focus on the organisational objectives (Hong et al. 2003).

ISSiO uses governance to provide boundaries and procedures for employees along with their roles and responsibilities and considers the organisation's risks and culture, performance and assurance, SETA, suppliers and customers (Brotby et al. 2006; Hinde 2002). Information security strategy is built on IT products and solutions but extends to include the employees in the business. Specifically ISSiO integrates director-level security sponsorship and hierarchical structures that provide security governance (Kayworth and Whitten 2010). ISSiO requires the attention and support of the board of directors and CEO because they are accountable for its outcomes. They affect ISSiO by using corporate governance, specifically a corporate information security policy, as a tool to communicate with and direct management in the organisation. Two-way communication is then required back from management to the board and executive in the form of regular progress reports (ISO/IEC 2013; McFadzean et al. 2006; Posthumus and von Solms 2004; Vroom and Von Solms 2004). ISSiO must consider corporate governance and provide those responsible for security with autonomy (Von Solms and Von Solms 2004).

ISSiO constituents include risk management components such as disaster recovery and business continuity, insurance, audits and new business units and groups (Cline and Jensen 2004). Without a focus on business continuity, it is entirely possible than in the event of an ICT infrastructure disaster a lack of business continuity translates directly into quantifiable revenue loss (Van Der Haar and Von Solms 2003).

Information security strategy needs to focus on people and process not tools, as these are the main causes of security failure (Da Veiga and Eloff 2010). ISSiO is preventative in nature and seeks to protect against rational individuals perpetrating attacks rather than automated technical attacks. The preventative approach relies heavily on deterrence and advocates that effectiveness is derived from sanctions being believed to be swift, severe and certain (D'Arcy and Herath 2011).

*Inter-organisational level*

The inter-organisational level of information security is where organisational benefits can potentially be mutually shared by contributing organisations for their individual success and factors that influence this are examined in the following section. At an inter-organisational level, compliance must be audited and a firm's auditing costs, incurred through engagement with an external auditor, can be lowered through a focus on IT assurance. This IT assurance includes high-quality IT documentation and an emphasis on systems security which lowers the cost because it makes the work of an auditor easier and quicker, therefore considerably lowering the time and materials auditing cost (Banker et al. 2010).

ISSiO facilitates information warfare, which forms just one layer of a conflict with an adversary. The four layers of a nation attack are political, which then escalates to economic sanctions, then information warfare and finally full kinetic warfare (Baskerville 2010). Some information assets may be resources that create strategic competitive advantage for organisations. If these lose their confidentiality through a security incident, then their integrity may be lost forever, along with the value of the advantage. When a security incident of this nature is disclosed to the market, there are implications for the organisation's share price (Campbell et al. 2003).

ISSiO is the process of dynamically assessing customer perceptions of the organisation's online transactions, with a view to increasing the security of transactions in order to prevent a decrease in brand trust in the marketplace. Regulatory pressures have increased the requirement for this defensive process (Datta and Chatterjee 2008). ISSiO must include an organisation's business and policy cyber considerations and depends on the political environment in an organisation's country of origin, which must synchronise with that of governments from other countries. The legal frameworks in various countries must harmonise globally to allow prosecution in the event of an attack. Shouldering the responsibility for lowering attacks will involve constitutional examination for potential conflicts, a willingness to collaborate and a system for measuring attacks however the benefits are that the world will be a safer place (Kim et al. 2012).

### 3.3.3 Yields

Yields are the goals achieved from the successful use of ISSiO and emerged as a theme in the information systems literature after conducting the thematic analysis described in Section 3.3. At an individual level, there were no apparent benefits arising from ISSiO, nor were there any apparent at a group level of analysis.

At an organisational level, the security goals are to ensure knowledge assets' confidentiality, integrity and availability (Ahmad et al. 2014a). another yield is that high quality information is made readily available (Doherty and Fulford 2006). Prevention of potential losses is an objective but depends on the volume of organisational information assets, business continuity capabilities, profitability, threat intelligence and risk appetite. Security budgets to achieve this prevention should be bounded by expected probable losses (Anderson and Choobineh 2008). Loss prevention efforts should also guard against revenue loss (Van Der Haar and Von Solms 2003). Performance reporting is another goal but requires tracking of key KPIs including systems, assigned assets, people, processes, compliance and auditing and customer service (Booker 2006). Finally, the protection of competitive advantage is an obvious goal (Cegielski et al. 2013).

At an inter-organisational level, ISSiO yields can include the misdirection of an adversary's attack assets, even from other nation-states, to protect information assets and physical critical infrastructure assets. Yields can also include the disablement of adversary CI, reduce foreign military abilities and impair foreign government operations (Baskerville 2010). ISSiO can also lower the risk of adverse litigation outcomes and achieve information confidentiality, integrity, availability, authenticity and non-repudiation (Brotby et al. 2006). An important benefit is share price protection (Campbell et al. 2003). Regulatory compliance avoids adverse sanctions by ensuring external agencies are kept fully informed (Banker et al. 2010). ISSiO yields also include retaining customers, security incident prevention, improved business processes and public reputation (Cline and Jensen 2004). Failure to implement an ISSiO sensibly may result in estranged customers and tarnished reputation (Datta and Chatterjee 2008; Oshri et al. 2007).

### 3.3.4 Key findings of thematic analysis

A number of gaps in knowledge have appeared through the conduct of this research. At an individual level of analysis, there appears to be very little research conducted into the role of an individual when supporting ISSiO. There appears to be many contributors to various aspects of the ISSiO construct but there does not seem to be any one unified conceptualisation or theory. Information security cannot be managed only at an organisational level but must include an inter-organisational level as well to take advantage of most of the yields.

Table 3 presents a thematic map of ISSiO derived from the results of the literature review, as described in the previous sections, and summarises the key themes found.

| **Antecedents** | **ISSiO Constituents** | **Yields** |
| --- | --- | --- |
| *Inter-organisational* | *Inter-organisational* | *Inter-organisational* |
| Regulatory compliance | Regulatory compliance | Foreign adversary impairment |
| Industrial and economic factors | Information warfare | Litigation risk management |
| Political and economic factors | Information asset protection | Share price protection |
| Political and legal factors | Environment scanning | Regulatory compliance |
| External threat environment |  | Public reputation |
| Standards |  | Customer trust |
| *Organisational* | *Organisational* | *Organisational* |
| Valuable information | Boardroom accountability | Confidentiality, integrity and availability |
|  | Quality improvement | Probably loss mitigation |
|  | Information asset management | Performance reporting |
|  | Labour source | Competitive advantage protection |
|  | Risk management |  |
|  | Organisational agility |  |
|  | Governance |  |
|  | Business continuity |  |
|  | People and process |  |
|  | Incident prevention |  |
|  | Policy |  |
| *Group* | *Group* | *Group* |
| Ubiquitous information availability | Knowledge leakage prevention | None |
|  | Security budget |  |
|  | Responsibility |  |
|  | Controls |  |
|  | Incident response |  |
|  | ICT infrastructure |  |
| *Individual* | *Individual* | *Individual* |
| None | None | None |

*Table 3. Thematic Map of Results from Literature Review of ISSiO*

## 4　CONCLUSION

This literature review illustrates various aspects of ISSiO and key themes were explored and grouped. Yet, there is no single, well-developed conceptualisation apparent in the literature that comprehensively explains the ISSiO construct and its relationships. Additionally, information security is ostensibly lacking to a large extent from the strategic organisational literature and even from strategic information systems literature. A paradigm shift is required to extend from internally-focussed protection of organisation-wide information towards a strategic view that considers the inter-organisational level. The following section offers suggestions to address these gaps through the conduct of future research, which could include positing a general framework to allow information systems researchers to investigate how ISSiO relates to inter-organisational strategy.

### 4.1　Contribution

Based on our review and a cumulative research tradition, we now construct a definition proposing the meaning of ISSiO:

> "Information security strategy is an organisational-wide framework of conceptual elements from individual up to inter-organisational level, which is informed by antecedent threat conditions in order to yield measurable information security benefits internal or external to the organisation."

### 4.2　Limitations of Research into Information Security Strategy

The ISSiO construct developed so far is potentially of great benefit to organisations seeking to adopt an overall strategy for their information security. We understand firstly, the precursor conditions which when met, cause organisations to consider the use of ISSiO; secondly, the constituent elements of an ISSiO for operationalization; and thirdly, the benefits that can be enjoyed by an organisation upon successful implementation. Given that, we still have limitations impeding our understanding of ISSiO. These are described in the next section.

Firstly, a significant amount of research conceptualises ISSiO as a plan, which identifies the construct as a static document, bereft of dynamic processes to ensure its validity when responding to immediate changes in the external environment. This gives rise to construct validity issues as having a plan is important, but not a precondition for an organisation to vary its ISSiO based on persistent incident detection and response (Straub et al. 2004).

Secondly, the information systems literature contains analysis on ISSiO from various levels within an organisation, largely focusing on the organisational perspective. This stratified perspective has its own properties and varies from an inter-organisational level, for example in terms of complexity and focus on external factors. Therefore, the nomological network of terms will be different for each level.

Thirdly, measurement issues arose in our study when we found that information systems researchers either did not adequately explain the dimensions with which to measure the elements of the ISSiO construct at each level or defined theoretical measures for one level and then operationalised them at another (Baskerville and Dhillon 2008). Additionally, tangible aspects of ISSiO such as the use of technical controls were perceived to be very measurable through reporting but intangible aspects such as employee attitudes towards security less so.

### 4.3　Future Research Directions

In addition to conducting further research on the gaps identified in this paper, there are several prospects for information systems researchers to develop the body of knowledge that currently exists on ISSiO. Answers to these questions have implications for practice. This study provides the impetus hopefully for future research into ISSiO, strategic information systems and organisational strategy.

Firstly, military strategy has influenced business management theory in many ways, most illustratively by the adaptation of the de-militarised zone (DMZ) concept by computer network architects. How can military strategy contribute to our understanding of ISSiO? What aspects of warfare, including embodying any supporting theory e.g. possibility theory, are pertinent to ISSiO?

Secondly, given the strong links from ISSiO to organisational strategic theory apparent in the literature, what lessons does business strategy have for ISSiO? How can ISSiO be integrated with business strategy? Is there a dependence on ISSiO to achieve organisational success, and if so, how is this success defined? What preconditions would prompt an organisation to strategically consider the use of ISSiO? Are there avenues to generate additional competitive advantage through ISSiO? Are there differences in ISSiO between public and private sectors?

Thirdly, information systems researchers could generate a framework or model to explain the phenomena that collectively form the ISSiO construct. What are the constituent elements of ISSiO and how do these relate to each other? How would ISSiO be operationalised within an organisation? To what extent will compliance culture influence the effectiveness of ISSiO operationalisation (Shedden et al. 2010; Tan et al. 2010)? How does ISSiO relate to strategic information systems? How does ISSiO relate to organisational strategy? What is the role of the individual level in ISSiO? How do levels of analysis apply in the digital realm?

Finally, there are a number of information systems scholars who have researched the theory underlying ISSiO, including for example deterrence, prevention, surveillance, detection, response, deception, perimeter defence, compartmentalisation and layering (Ahmad et al. 2014b; Beebe and Rao 2009; D'Arcy and Herath 2011). What would further analysis of these theories reveal about ISSiO? What does systems theory have to offer ISSiO?

## 5   REFERENCES

Ahmad, A., Hadgkiss, J., and Ruighaver, A.B. 2012. "Incident Response Teams–Challenges in Supporting the Organisational Security Function," Computers & Security (31:5), pp 643-652.

Ahmad, A., Bosua, R., and Scheepers, R. 2014a. "Protecting Organizational Competitive Advantage: A Knowledge Leakage Perspective," Computers & Security (42), pp 27-39.

Ahmad, A., Maynard, S.B., and Park, S. 2014b. "Information Security Strategies: Towards an Organizational Multi-Strategy Perspective," Journal of Intelligent Manufacturing (25:2), pp 357-370.

Anderson, E.E., and Choobineh, J. 2008. "Enterprise Information Security Strategies," Computers & Security (27:1), pp 22-29.

Backhouse, J., and Dhillon, G. 1996. "Structures of Responsibility and Security of Information Systems," European Journal of Information Systems (5:1), pp 2-9.

Baets, W. 1992. "Aligning Information Systems with Business Strategy," Journal of Strategic Information Systems (1:4), pp 205-213.

Banker, R., Chang, H., and Kao, Y.-C. 2010. "Evaluating Cross-Organizational Impacts of Information Technology – an Empirical Analysis," European Journal of Information Systems (19:2), pp 153-167.

Baskerville, R. 2010. "Third-Degree Conflicts: Information Warfare," European Journal of Information Systems (19:1), pp 1-4.

Baskerville, R., and Dhillon, G. 2008. "Information Systems Security Strategy: A Process View," in: Information Security: Policy, Processes, and Practices. Advances in Management Information Systems, D.W. Straub, S.E. Goodman and R. Baskerville (eds.). Armonk, NY: M. E. Sharpe., pp. 15-45.

Baskerville, R., Spagnoletti, P., and Kim, J. 2014. "Incident-Centered Information Security: Managing a Strategic Balance between Prevention and Response," Information & Management (51:1), pp 138-151.

Beebe, N.L., and Rao, V.S. 2009. "Examination of Organizational Information Security Strategy: A Pilot Study," AMCIS 2009 Proceedings.

Beebe, N.L., and Rao, V.S. 2010. "Improving Organizational Information Security Strategy Via Meso-Level Application of Situational Crime Prevention to the Risk Management Process," Communications of the Association for Information Systems (26:17), pp 329-358.

Booker, R. 2006. "Re-Engineering Enterprise Security," Computers & Security (25:1), pp 13-17.

Bowen, P., Hash, J., and Wilson, M. 2006. Sp 800-100. Information Security Handbook: A Guide for Managers.

Brotby, W., Bayuk, J., and Coleman, C. 2006. Information Security Governance: Guidance for Boards of Directors and Executive Management. Illinois, IT Governance Institute.

Burnburg, M.K. 2003. "A Proposed Framework for Business Information Security Based on the Concept of Defense-in-Depth." Springfield: University of Illinois.

Campbell, K., Gordon, L.A., Loeb, M.P., and Zhou, L. 2003. "The Economic Cost of Publicly Announced Information Security Breaches: Empirical Evidence from the Stock Market," Journal of Computer Security (11:3), pp 431-448.

Cegielski, C.G., Bourrie, D.M., and Hazen, B.T. 2013. "Evaluating Adoption of Emerging It for Corporate It Strategy: Developing a Model Using a Qualitative Method," Information Systems Management (30:3), pp 235-249.

Cline, M., and Jensen, B. 2004. "Information Security: An Organizational Change Perspective," AMCIS 2004 Proceedings.

D'Arcy, J., and Herath, T. 2011. "A Review and Analysis of Deterrence Theory in the Is Security Literature: Making Sense of the Disparate Findings," European Journal of Information Systems (20:6), pp 643-658.

Da Veiga, A., and Eloff, J.H.P. 2007. "An Information Security Governance Framework," Information Systems Management (24:4), pp 361-372.

Da Veiga, A., and Eloff, J.H.P. 2010. "A Framework and Assessment Instrument for Information Security Culture," Computers & Security (29:2), pp 196-207.

Datta, P., and Chatterjee, S. 2008. "The Economics and Psychology of Consumer Trust in Intermediaries in Electronic Markets: The Em-Trust Framework," European Journal of Information Systems (17:1), pp 12-28.

Doherty, N.F., and Fulford, H. 2006. "Aligning the Information Security Policy with the Strategic Information Systems Plan," Computers & Security (25:1), pp 55-63.

Flores, W.R., Antonsen, E., and Ekstedt, M. 2014. "Information Security Knowledge Sharing in Organizations: Investigating the Effect of Behavioral Information Security Governance and National Culture," Computers & Security (43), pp 90-110.

Hinde, S. 2002. "Security Surveys Spring Crop," Computers & Security (21:4), pp 310-321.

Hong, K.-S., Chi, Y.-P., Chao, L., and Tang, J.-H. 2003. "An Integrated System Theory of Information Security Management," Information Management & Computer Security (11:5), pp 243-248.

ISO/IEC. 2013. "Iso/Iec 27014:2013 Information Technology — Security Techniques — Governance of Information Security." Geneva, Switzerland: ISO/IEC.

Johnson, M.E., and Goetz, E. 2007. "Embedding Information Security into the Organization,"  (3), pp 16-24.

Kayworth, T., and Whitten, D. 2010. "Effective Information Security Requires a Balance of Social and Technology Factors," MIS Quarterly Executive (9:3), pp 163-175.

Kelly, B.J. 1999. "Preserve, Protect, and Defend," The Journal of Business Strategy (20:5), pp 22-25.

Kim, S.H., Wang, Q.-H., and Ullrich, J.B. 2012. "A Comparative Study of Cyberattacks," Communications of the ACM (55:3), p 66.

Leidner, D.E., and Kayworth, T. 2006. "Review: A Review of Culture in Information Systems Research: Toward a Theory of Information Technology Culture Conflict," MIS Quarterly (30:2), pp 357-399.

McFadzean, E., Ezingeard, J.-N., and Birchall, D. 2006. "Anchoring Information Security Governance Research: Sociological Groundings and Future Directions," Journal of Information System Security (2:3), pp 3-48.

Oshri, I., Kotlarsky, J., and Hirsch, C. 2007. "Information Security in Networkable Windows-Based Operating System Devices: Challenges and Solutions," Computers & Security (26:2), pp 177-182.

Park, S., and Ruighaver, T. 2008. "Strategic Approach to Information Security in Organizations," ICISS. International Conference on Information Science and Security, 2008: IEEE, pp. 26-31.

Posthumus, S., and von Solms, R. 2004. "A Framework for the Governance of Information Security," Computers & Security (23:8), pp 638-646.

Roberts, N., Galluch, P.S., Dinger, M., and Grover, V. 2012. "Absorptive Capacity and Information Systems Research: Review, Synthesis, and Directions for Future Research," MIS Quarterly (36:2), pp 625-648.

Ryan, J.J., and Ryan, D.J. 2006. "Expected Benefits of Information Security Investments," Computers & Security (25:8), pp 579-588.

Shedden, P., Ruighaver, T., and Ahmad, A. 2010. "Risk Management Standards – the Perception of Ease of Use," Journal of Information Systems Security (6:3), pp 23-41.

Straub, D., Boudreau, M.-C., and Gefen, D. 2004. "Validation Guidelines for Is Positivist Research," The Communications of the Association for Information Systems (13:1), p 63.

Sveen, F., Torres, J., and Sarriegi, J. 2009. "Blind Information Security Strategy," International Journal of Critical Infrastructure Protection (2:3), pp 95-109.

Tan, T., Ruighaver, A.B., and Ahmad, A. 2010. "Information Security Governance: When Compliance Becomes More Important Than Security," The IFIP TC-11 24th International Information Security Conference Brisbane, Australia: Springer, pp. 55-67.

Taylor, R.G., and Robinson, S.L. 2014. "The Roles of Positive and Negative Exemplars in Information Security Strategy," Academy of Information and Management Sciences Journal (17:2), pp 57-79.

Tutton, J. 2010. "Incident Response and Compliance: A Case Study of the Recent Attacks," Information Security Technical Report (15:4), pp 145-149.

Van Der Haar, H., and Von Solms, R. 2003. "A Model for Deriving Information Security Control Attribute Profiles," Computers & Security (22:3), pp 233-244.

Van Niekerk, J.F., and Von Solms, R. 2010. "Information Security Culture: A Management Perspective," Computers & Security (29:4), pp 476-486.

Von Solms, B., and Von Solms, R. 2004. "The 10 Deadly Sins of Information Security Management," Computers & Security (23:5), pp 371-376.

Vroom, C., and Von Solms, R. 2004. "Towards Information Security Behavioural Compliance," Computers & Security (23:3), pp 191-198.

Webster, J., and Watson, R.T. 2002. "Analyzing the Past to Prepare for the Future: Writing a Literature Review," Management Information Systems Quarterly (26:2), pp xiii-xxiii.

## ACKNOWLEDGEMENTS

The authors would like to thank the reviewers for their valuable contributions to this paper.

## COPYRIGHT